\documentclass[manuscript,screen]{acmart}
\setcopyright{none}
\usepackage{multirow}
\usepackage{amsmath}
\usepackage{enumerate}

\usepackage{float}
\usepackage{graphicx}
\usepackage{rotating}

\usepackage{hyperref}
\usepackage{url}
\usepackage{makecell}
\newcommand{\BRANCH}{Branches}
\AtBeginDocument{%
  \providecommand\BibTeX{{%
    \normalfont B\kern-0.5em{\scshape i\kern-0.25em b}\kern-0.8em\TeX}}}

\acmConference[]{ }{ }{ }
\acmPrice{ }
\acmISBN{ }

\begin{document}

\title[Fuzzers for stateful systems]{Fuzzers for stateful systems: Survey and Research Directions}

\author{Cristian Daniele}
\orcid{1234-5678-9012}
\email{cristian.daniele@ru.nl}
\author{Seyed Behnam Andarzian}
\email{seyedbehnam.andarzian@ru.nl}
\orcid{1234-5678-9012}
\author{Erik Poll}
\orcid{1234-5678-9012}
\email{erikpoll@cs.ru.nl}
\affiliation{%
  \institution{Radboud University}
  \streetaddress{P.O. Box 1212}
  \city{Nijmegen}
  \state{}
  \country{The Netherlands}
  \postcode{43017-6221}
}


\begin{abstract}
Fuzzing is a security testing methodology effective in finding bugs.
In a nutshell, a fuzzer sends multiple slightly malformed messages to the software under test, hoping for crashes or weird system behaviour.
The methodology is relatively simple, although applications that keep internal states are challenging to fuzz.
The research community has responded to this challenge by developing fuzzers tailored to stateful systems, but a clear understanding of the variety of strategies is still missing.
In this paper, we present the first taxonomy of fuzzers for stateful systems and provide a systematic comparison and classification of these fuzzers.
\end{abstract}





\maketitle

\section{Introduction}

With fuzzing (or fuzz testing) a system is fed a large collection of
automatically generated inputs to find security vulnerabilities, in
particular memory corruption bugs.  Fuzzing is a great way to improve
software security: it can find lots of bugs with relatively little
effort.
The idea of fuzzing goes back to the late 1980s \cite{Miller1990} but
interest in fuzzing exploded in the 2000s. Major game changers here
have been the advent of \emph{white-box fuzzing} using symbolic (or
more precisely, concolic) execution in the SAGE fuzzer \cite{SAGE2008}
and the advent of \emph{grey-box fuzzing}, also known as
\emph{evolutionary fuzzing}, pioneered in the fuzzer AFL \cite{afl},
where code execution paths are monitored to guide the generation of
inputs.  Both approaches avoid the need of having the user provide an
explicit grammar describing the input format.

Traditionally, most fuzzers target \emph{stateless} systems, where the system
under test takes a single input, say a JPEG image; the fuzzer then tries many
possible inputs, including many malformed ones.  This survey focuses on the
fuzzing of \emph{stateful systems}. By a stateful system, we mean a system
that takes a \emph{sequence} of messages as input, producing outputs along the
way, and where each input may result in an internal state change.  Most
protocols, including most network protocols, are stateful. So when people talk
about `network fuzzing' \cite{Helme2021} or `network protocol fuzzing'
\cite{Gorbunov2010,GANFuzz2018}, they are usually talking about the fuzzing of
stateful systems.  Obviously, fuzzing stateful systems is harder than fuzzing 
stateless systems, as the internal state changes increase the state space that
we try to explore. Moreover, it may be hard for a fuzzer to reach
`deeper' states. Indeed, fuzzing stateful systems is listed as one of the
challenges in fuzzing by Boehme et al.\ \cite{Boehme2021}.

There are some good survey papers about fuzzing, e.g.\ \cite{Manes2021,Zhu2022},
but these do not investigate the issues of fuzzing stateful systems in any
depth, even though these surveys do include some fuzzers for stateful systems.
There is a growing number of fuzzers specifically for stateful systems, which
raises questions about the approaches these fuzzers take, their commonalities
and differences, and their pros and cons, which this paper tries to answer.

The outline of this paper is as follows.  Section~\ref{sec:terminology}
defines some basic concepts and terminology for discussing stateful
systems.  Section~\ref{sec:Background} presents the traditional
classification for fuzzers used in the literature and discusses some
of the differences between fuzzing stateless and stateful systems.
Section~\ref{sec:stateful_fuzzer_fuzzers} presents our taxonomy for
fuzzers for stateful systems, classifies existing fuzzers and compares
the various approaches.

\section{Concepts and Terminology}\label{sec:terminology}

By a stateful system we mean a system that takes a sequence of
messages as input, producing outputs along the way, and where each
input may result in an internal state change.  To avoid confusion, we
reserve the term \emph{message} or \emph{input message} for the
individual input that the System Under Test (SUT) consumes at each
step and the term \emph{trace} for a \emph{sequence} of such messages
that make up the entire input.

We use the term \emph{response} for the output that the SUT produces
along the way. In the case of a synchronous protocol, there is usually
one response after each input message. In this case, the state machine
describing the input-output behaviour will be a Mealy machine.

The input language of a stateful system consists of two
levels\footnote{For text-based protocols, as opposed to binary
protocols, there may even be a third level, namely the character set
or character encoding used, but none of the fuzzers studied use
that.}: 1) the language of the individual messages, which we will
refer to as the \emph{message format}, and 2) the language of traces,
built on top of that.  A description or specification of such an input
language will usually come in two parts, one for each of the levels:
for example, a context-free grammar for the message format and
a finite state machine describing sequences of these messages.  We
will call the latter the \emph{state model} or, if it is described as
a state machine, the \emph{protocol state machine}.

The state model usually involves a notion of \emph{message type} where
messages of one type trigger a different transition than messages of
another type.  So then the set of messages types is the input alphabet
of any protocol state machine. This alphabet will typically abstract
away from payloads inside messages.  For some protocols, messages
simply include an instruction byte in a header that determines the
message type.  For output messages, it is common to distinguish
between error responses and non-error responses, and possibly a
finer-grained distinction of error responses based on the error code.
Fuzzers that try to infer the protocol state machine (using active or
passive learning, as will be discussed later) may require the user to
specify the abstraction function that maps concrete messages to
message types or even to provide an implementation of this function.
There can be two abstraction functions, one for input messages and one
for responses.

In some protocols the format of input messages and responses is very
similar. For fuzzing it is then a good strategy to also include
responses as inputs as this may trigger unexpected behaviour: client
and server are likely to share a part of the codebase and one may then
accidentally will process messages only intended for the
other\footnote{CVE-2018-10933 is an interesting example of a bug of
this kind in Libssh: if the message that the server sends to a client
to confirm that the client has successfully authenticated is sent
\emph{to} the server, a malicious client could skip the whole
authentication phase.}

\smallskip

We use the term \emph{protocol state} to refer to the abstract state of an SUT
that determines how future input messages are handled.  The SUT will have a
concrete program state, which is related to this protocol state, but which
usually carries much more detail.

The term `state' can quickly become overloaded: not only the SUT has state, even
if it implements a stateless protocol, but the fuzzer itself also has a state.
We use the term \emph{stateful fuzzing} to refer to the fuzzing of stateful
systems, but we avoid the term `stateful fuzzer' as even a fuzzer for stateless
systems will have internal state.

There are basically two ways for an SUT to record the protocol state:
1) it can keep track of the protocol state using program variables
that record state information or 2) the state can be more implicitly
recorded by the program point where the execution is at (and possibly
the call stack). Of course, these two ways can be combined.  For
fuzzers that use a grey- or white-box approach (discussed in more
detail later), the difference can be important:  white-box approaches
that observe the values of program variables will work better for 1)
than for 2), whereas grey-box approaches that observe execution paths
will work better for 2) than for 1).


\paragraph{Stateless vs stateful systems}

There is not always a clear border between stateless systems and
stateful systems. For example, if a system has a memory leak then it
is (unintentionally) stateful, even though its behaviour will appear
to be stateless for a very long time, namely until it runs out of
memory and crashes.

More generally, we can think of any stateful system that takes a
sequence of messages as input as a stateless system which takes that
whole trace as single input.  Conversely, we can view a stateless
system that takes a single complex value as input as a stateful system
that processes a sequence of smaller inputs, say bits or bytes. For
instance, a stateless program that processes JPEGs, which will always
process the same JPG in the same way, can be viewed as a
\emph{stateful} program that takes a sequence of bytes as input, and
which will process the same byte in different ways depending on where
in the JPEG image it occurs.  But a fundamental difference between a
stateful system and a stateless system viewed as stateful one in this
way is that the former will typically provide some observable output
after processing each input.  Another difference is knowing that
inputs are made up of smaller messages can help in making useful
mutations, by swapping the order of messages or repeating messages.

\smallskip

Some stateless systems can process sequences of inputs like stateful
systems do, but then the idea is that previous inputs do not have any
effect on how subsequent inputs are handled.  Some fuzzers use this
possibility to avoid the overhead of having to restart the SUT between
inputs. This is called \emph{persistent fuzzing}.  This does
then involve a sequence of inputs, but it is the polar opposite of
stateful fuzzing: the goal is not to explore the statefulness of the
SUT, but rather it presupposes that there is no statefulness.

%

\section{Background}
\label{sec:Background}

This section discusses existing classifications of fuzzers used in the
literature, as this provides a starting point for our classification
of stateful fuzzing, and makes some initial observations about how and
why fuzzing stateful systems is different.

There are some good surveys papers about fuzzing, but none pay much
attention to issues specific to stateful fuzzing. 
The survey by Zhu et al.\ \cite{Zhu2022} classifies close to 40
fuzzers. Only two of these, namely AFLNet \cite{aflnet} and de Ruiter et al.\
\cite{deRuiter2015} specifically target stateful systems. 
The more extensive survey by Manes et al.\ \cite{Manes2021} categorises
over 60 fuzzers. Thirteen of these are fuzzers for `network protocols'
so presumably these are fuzzers geared to fuzzing stateful systems;
all of these 13 fuzzers are black-box fuzzers.  One section in this
survey (Section~5.1.2) discusses the statefulness of the SUT: here it discusses
the inference of state machine models.


The only attempt at classifying fuzzers for stateful systems that we
are aware of is given in the paper by Yu et al.\ about SGPFuzzer
\cite{SGPFuzzer}: Table~1 in this paper lists twelve other fuzzers for
stateful systems (or "protocol fuzzers" in the terminology used in the
paper), namely AutoFuzz \cite{Gorbunov2010}, AspFuzz, SECFuzz
\cite{tsankov2012secfuzz}, Sulley
\footnote{https://github.com/OpenRCE/sulley}, BooFuzz
\cite{pereyda2019boofuzz}, Peach
\footnote{https://wiki.mozilla.org/Security/Fuzzing/Peach}, SNOOZE
\cite{banks2006}, PULSAR \cite{pulsar}, TLS-fuzzer, DTLS-fuzzer,
ICS-fuzzer, and NLP-fuzzer.  The authors identify four challenges
based on the shortcomings of these 12 fuzzers and then design
SGPFuzzer to address these.  Unfortunately, the definitions of the
features used for the comparison are left implicit and the comparison
fails to point out some very fundamental differences between tools,
for instance that the man-in-the-middle nature of AutoFuzz and SecFuzz
comes with an important inherent limitation, namely that the order of
messages cannot be fuzzed (as we discuss in
Section~\ref{sec:man_in_the_middle_fuzzers}). 

\subsection{Existing classifications of fuzzers} \label{sec:stateless}
The standard classification of fuzzers in the literature (e.g.\
\cite{Manes2021, Boehme2021, Zhu2022}) distinguishes \emph{black-box},
\emph{grey-box} and \emph{white-box fuzzers}, where the black-box
fuzzers are sub-divided into \emph{grammar-based} fuzzers and
\emph{mutation-based fuzzers}.  Even though this classification is
fairly standard, the terminology varies between papers, and there are
combinations of approaches that do not neatly fit into one of these
categories.  We discuss this classification in more detail below.

\paragraph{Black-box fuzzers} As the name suggest, black-box fuzzers
only observe the SUT from the outside. To stand any change of
producing interesting inputs, black-box fuzzers require some knowledge
of the input format. 
  One approach here, taken by \emph{generation-based} aka
    \emph{grammar-based} fuzzers, is that the fuzzer is given
    knowledge of the input format in the form of grammar or model. A
    downside of such fuzzers is the effort required to produce such a
    model or grammar.  
  Another approach, taken by \emph{mutation-based fuzzers}, is
    that the fuzzer is supplied with a set of sample inputs which are
    then mutated in the hope of discovering bugs.

Most grammar-based fuzzers allow users to supply grammar for an
arbitrary input format (or protocol) they want to fuzz, but there are
also grammar-based fuzzers which have a specific input format
hard-coded in them, such as the KiF fuzzer \cite{KiF2007} for the SIP
protocol.  There are also commercial fuzzers for fuzzing specific
protocols, for example, Codenomicon's DEFENSIS fuzzer
\footnote{www.codenomicon.com/defensics/} (since acquired by
Synopsys), which grew out of the PROTOS \cite{kaksonen2001software}
project at the University of Oulu in Finland started in 1999.

Some fuzzers combine the generation-based and mutation-based approach.  A
grammar-based fuzzer should not only produce grammatically correct inputs, but
also malformed ones; this either has to involve some form of mutation or the
grammar has to be too `loose' to begin with.  Conversely, a mutation-based
fuzzer can be given some knowledge about the input format, for instance by
providing a list of keywords or byte values with a special meaning, which the
fuzzer can use in mutations in the hope of generating interesting mutations.

\paragraph{White-box fuzzers}
White-box fuzzers require access to the (source or binary) program code and
analyse that code in order to provide interesting inputs. With access
to the code, it is possible to see which branches there are to then
construct inputs that trigger particular branches. Typically
white-box fuzzers use symbolic or concolic execution to construct
interesting test cases. Microsoft's SAGE fuzzer \cite{SAGE2008} is the best-known example of this class.

\paragraph{Grey-box fuzzers}

Grey-box fuzzers occupy the middle ground and can observe some aspects
of the SUT as it executes and use this feedback to steer the fuzzer.
This is also called \emph{evolutionary fuzzing} as the inputs will
gradually evolve into more interesting mutations.
Grey-box fuzzers can be considered as a special kind of mutational
fuzzers because the evolution always involves mutation.
Grey-box fuzzers are sometimes called smart mutational fuzzers; the
black-box mutational fuzzers that lack a feedback mechanism to guide
the evolution are then called dumb mutational fuzzers.

Grey-box fuzzers often require some instrumentation of the code or
running the code in some emulator. The approach has been popularised
by the fuzzer AFL, which observes the code execution path -- or, more
precisely, the branches are taken in the execution -- to see if some
input mutation results in new execution paths.  Grey-box fuzzers that
observe the execution path in this way are also called
\emph{coverage-guided greybox fuzzers (CGF)}.  This approach has
proved to be very successful, providing much better coverage than
`dumb' mutational fuzzers but without the work of having to provide a
grammar.  

All fuzzers that involve mutation -- dumb mutational fuzzers,
evolutionary fuzzers, but also grammar-based fuzzers that use mutation
-- can be parameterised by \emph{mutation primitives}, for instance
random bit flips, repeating sub-sequences of the input, or inserting
specific bytes, characters, or keywords.

\paragraph{Alternative classifications}

Instead of classifying fuzzers into white-, grey- and black-box, an orthogonal
classification is to consider the kind of applications targeted and the kind of
input this involves \cite{Manes2021}: e.g.\ some fuzzers are geared towards
fuzzing file formats, others to network traffic, and others still to web
applications or OS kernels.  Fuzzers for web applications are often called
`scanners'.  There is a relation between this classification and statefulness:
applications that take a file as input are usually not stateful, whereas
applications that implement a network protocol usually are.

\subsection{White-, grey-, and black-box fuzzing for stateful systems}\label{sec:wgbfuzzing}

For stateful systems some basic observations about the classification
into white-, grey-, and black-box can be made:

\begin{itemize}

\item The terms `grey-box fuzzing' and `evolutionary fuzzing' are often
 used as synonyms, but for stateful systems, they are not:
 for a stateful system the evolution of inputs can also be
 steered by the outputs of the SUT, which is then evolutionary but
 black-box. This is a key difference between a stateful
 and a stateless system: the response that the SUT produces is an
 observation that the fuzzer can make without any instrumentation
 of the code.

\item For grammar-based fuzzers it does not really matter if the
 SUT is stateful or not: the grammar describing the system can
 describe both the message format and the protocol state machine.

\item For dumb mutational black-box fuzzers it also does not
     matter that much if the SUT is stateful or not. Of course,
     it helps if fuzzer is aware of the
     fact that inputs are traces of messages, so that it can 
     try swapping the order of messages, removing messages or
     repeating messages as interesting mutations.  The same goes for
     any grammar-based fuzzer, which should also try re-ordering,
     repeating or dropping messages as interesting corruptions of the
     grammar.

\item The techniques used in grey-box and white-box fuzzing to observe
     program execution may shed some light on the state that the SUT
     is in.  But as discussed in Section~\ref{sec:terminology}, there
     are different ways in which the SUT can record protocol state:
     the state can be recorded in some program variables, it can be
     recorded in the program point that the SUT is in (and
     possibly the call stack), or a combination of
     these. The way in which the SUT does this can make a difference
     in how well some grey- or white-box technique can observe the
     protocol state.

 The statefulness of the SUT may complicate observation for a grey- or
    whitebox fuzzer.  For white-box fuzzers that rely on symbolic or
    concolic execution input, the statefulness of the SUT is obviously
    a serious complication: a symbolic execution of a program
    handling a single input can already be tricky, and the 
    execution for sequence of symbolic inputs will be an order of
    magnitude harder.  For example, if the SUT implements some loop to
    handle incoming messages then that loop would have to be unwound.


\end{itemize}

\subsection{Bug detection for stateful systems}

In addition to a mechanism to generate inputs, a fuzzer also requires some
mechanism to observe the SUT to detect if an input triggered a bug.  Typically
fuzzers look for memory corruption bugs that crash the SUT using sanitisers such
as ASan (AddressSanitizer) \cite{serebryany2012addresssanitizer}, MSan
(MemorySanitizer) \cite{MSan}, or older less efficient sanitisers such as
Valgrind.  When fuzzing programs written in memory-safe languages, e.g.\ when
fuzzing Java programs with Kelinci \cite{kersten2017poster}, instead of looking
for memory corruption bugs we can look for uncaught runtime exceptions; even if
these bugs cannot be exploited in the way memory corruption can, they can still
lead to Denial of Service problems.

A type of bug that is specific to stateful systems are deviations from
the correct state behaviour: if a system is expected to behave in a
specific way, for instance by only responding to certain inputs after
authentication, then a deviation from this behaviour may be a security
issue. For security protocols such as TLS any deviations from the
expected state behaviour are highly suspicious: security protocols are
very fragile and even small deviations may break the security
guarantees the protocol aims to provide.  Unlike the detection of
memory corruption bugs, this cannot be totally automated: it either
requires a specification of expected behaviour (or, conversely, of
unwanted behaviour), for instance with a state machine or in temporal
logic, or it requires some post-hoc manual analysis of the state
behaviour inferred by the fuzzer.

\section{Fuzzers for Stateful Systems}\label{sec:stateful_fuzzer_fuzzers}

\begin{table}
  \centering
  \begin{tabular}{|l|l|l|c|l|}
    \hline
    \textbf{Section} & \textbf{Category} & \textbf{Input required} & \textbf{\makecell[l]{Generate\\state model}} & \textbf{\makecell[l]{Human\\interaction}}\\
    \hline \ref{sec:grammar_based_fuzzers} &
     Grammar-based  & Grammar & No & No \\
    \hline \ref{sec:grammar_learner_fuzzers} &
    Grammar-learner & Sample traces & Yes & Yes / No \\
    \hline \ref{sec:evolutionary_fuzzers} &
    Evolutionary &  Sample traces & No & Yes / No \\
    \hline \ref{sec:evolutionary_grammar_based_fuzzers} &
    Evolutionary grammar-based & Grammar & No & No \\
    \hline \ref{sec:evolutionary_grammar_learner_fuzzers} &
    Evolutionary grammar-learner & Sample traces & Yes & No \\
    \hline \ref{sec:man_in_the_middle_fuzzers} &
    Man-in-the-Middle fuzzers & Live traffic & No & Yes / No \\
    \hline \ref{sec:machine_learning_fuzzers} &
    Machine learning fuzzers & Many sample traces & No & No \\
    \hline
  \end{tabular}
  \caption{The seven categories of fuzzers with their main
  characteristics. Human interaction refers to manual code or grammar annotation.}
  \label{table:categories}
\end{table}
  
\begin{figure}
  \centering
   \includegraphics[scale=0.45]{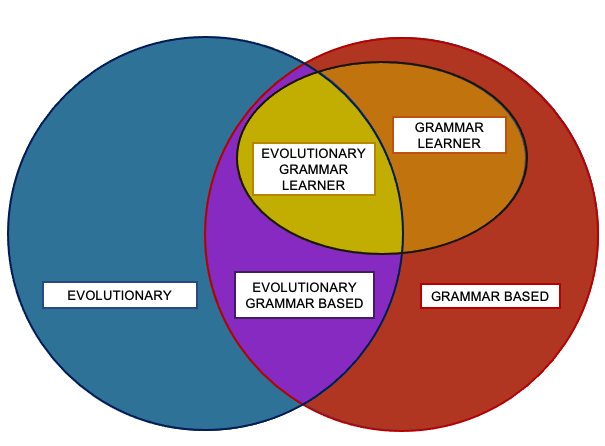}
   \caption{The five categories of fuzzers that involve a grammar, an
    evolutionary feedback mechanism, or both.}
    \label{fig:venn}
\end{figure}

We have identified seven categories of fuzzers for stateful fuzzing.
Table~\ref{table:categories} summarises the main characteristics of
each category.  Some categories can be regarded as a combination 
or sub-category of other categories, as illustrated in
Fig.~\ref{fig:venn}.  Before we discuss each category in more detail
in the sections below, we first discuss common ingredients
involved in some of them:
\begin{itemize}

\item Some fuzzers require sample traces as inputs, either a few traces to act
     as seeds to further mutation, or many traces so that grammar can be
     inferred or machine model can be trained. 

\item Many fuzzers involve some form of grammar.  This can be a
     grammar describing just for the message format, a grammar
     describing just the protocol state machine, or both.  Some
     fuzzers require such grammars as inputs, but others can provide 
     grammars that are inferred during the fuzzing as output.


\item Many fuzzers use some form of learning to infer information
     about the message format, the protocol state machine, or both.
     Evolution can be regarded as a form of learning because it
     produces and uses new knowledge about the input format, even
     though this knowledge is (usually) not expressed in the form of a
     regular expression, state machine, or context-fee grammar.

     Evolution is a form of \emph{active learning} because it
     involves interaction with the SUT, where the next input we try
     can depend on the outcome of previous tests.
     Some fuzzers use forms of \emph{passive learning} instead of (or in
     addition to) such active learning. By this we mean approaches
     where information about the input format is inferred \emph{after}
     a set of traces has been collected, so without interactively
     trying new experiments. 

     There is a long line of research into algorithms for inferring
     formal language descriptions, either actively or passively, which
     includes research into \emph{regular inference} and
     \emph{grammatical inference} that focus specifically on inference
     of regular expressions and context-free grammar, respectively.
     Research in this field is presented at the bi-annual
     International Conference on Grammatical Inference (ICGI) and
     there are entire textbooks on the subject (e.g.\
     \cite{delaHigueraDela201}).  For active learning of a protocol
     state machine, an algorithm that can be used is L*
     \cite{Angluin87} or one of its improvements, e.g.\ the TTT
     algorithm used in LearnLib \cite{isberner2014ttt}.  For the
     passive learning of protocol state machines, some fuzzers use
     ad-hoc solutions.  For instance, the fuzzer by Hsu et al.\
     \cite{hsu2008model} uses an algorithm called partial finite state
     automaton reduction.  An important limitation of some learning
     algorithms, notably L* and its improvements, is that they cannot
     deal with the non-deterministic behaviour of the SUT, as it will
     cause the algorithm to diverge.

     A very different form of (passive) learning used by some fuzzers
     is \emph{machine learning}. This does not produce knowledge in a
     nice concrete format like a regular expression, finite state
     machine, or context-free grammar. Also, it typically requires
     more samples. Still, possible advantages are that there are many
     existing machine learning approaches that can be used and that
     these may cope more easily with non-deterministic behaviour
     of the SUT.
\end{itemize}
Below we first give a general description of the seven categories of
fuzzers. In the subsequent sections, we discuss each category in more 
detail:

\begin{enumerate}

 \item \textbf{\textit{Grammar-based fuzzers}} Any grammar-based
      fuzzer can be used to fuzz stateful systems without any special
      adaptions.  The grammar that is supplied to the fuzzer will have
      to describe the two levels of the input language, with some
      rules of the grammar describing the message format and some
      rules describing the protocol state machine. Apart from that, no
      change in the fuzzer itself is needed, except that course,
      swapping, dropping and repeating messages are useful -- if not
      essential -- mutation strategies for the fuzzer to include. But
      for a stateless SUT where the format of the inputs is quite
      complex it can also be useful to include swapping, dropping and
      repeating sub-components of inputs as mutation strategies.

 \item \textbf{\textit{Grammar-learner fuzzers}} Whereas the
      grammar-based fuzzers require the users to provide a grammar,
      these fuzzers are able to extract a grammar from a set
      of sample traces.
    They can be considered as the sequential composition of two
    tools: a \textit{grammar extractor} that infers the grammar from
    a set of sample traces (using so-called passive grammatical inference)
    and a grammar-based fuzzer that then does the actual fuzzing using
    this inferred grammar. 

    As for the grammar-based fuzzers, for the grammar-learner fuzzers
    the statefulness of the SUT does not make any fundamental
    difference: it only means that the grammar will have two levels.
    So grammar-learner fuzzers can be applied to stateless as well as
    stateful SUTs.

\item \textbf{\textit{Evolutionary fuzzers}} These fuzzers basically
     take the same approach as stateless evolutionary fuzzers such as
     AFL:  they take some sample traces as
     initial input and mutate these using a feedback system to steer
     the mutation process.
     Of course, evolutionary fuzzers for stateful systems should be aware
     that an input trace is a sequence of messages and should include
     swapping, omitting or repeating these messages as 
     mutation strategies.

     A difference between stateful and stateless systems when it
     comes to evolutionary approaches of fuzzing is that the responses
     that a stateful SUT provides after individual messages can be
     used in the feedback to guide the evolution, as mentioned before
     in Section~\ref{sec:stateless}.

 \item \textbf{\textit{Evolutionary grammar-based fuzzers}} These
      fuzzers use both a grammar provided to the user to generate (correct,
      protocol-compliant) traces and an evolution mechanism to mutate
      these traces.
     We can think of them as evolutionary fuzzers that use a grammar
    instead of a set of sample input traces to provide
    the initial traces that will be mutated.  We can also think
    of them as grammar-based fuzzers that include a feedback mechanism
    to steer the evolution of mutations.  So in
    Fig.~\ref{fig:venn} they are the intersection of the
    evolutionary fuzzers and the grammar-based fuzzers.

\item \textbf{\textit{Evolutionary grammar-learner fuzzers}} This is
     the most complex category of fuzzers. These tools all use some
     form of grammar to describe the protocol state machine; one also
     uses a grammar to describe the message format.  They involve two
     feedback mechanisms to steer two forms of evolution: (i) one for the
     mutation of individual messages, in the style of conventional
     evolutionary fuzzers like AFL, and (ii) another for the mutation of
     sequences, which then infers a protocol state machine.  The
     second form of evolution is based on the response that the SUT
     provides as feedback, so it is black-box.

\end{enumerate}

\begin{figure}
\centering
 \includegraphics[scale=0.25]{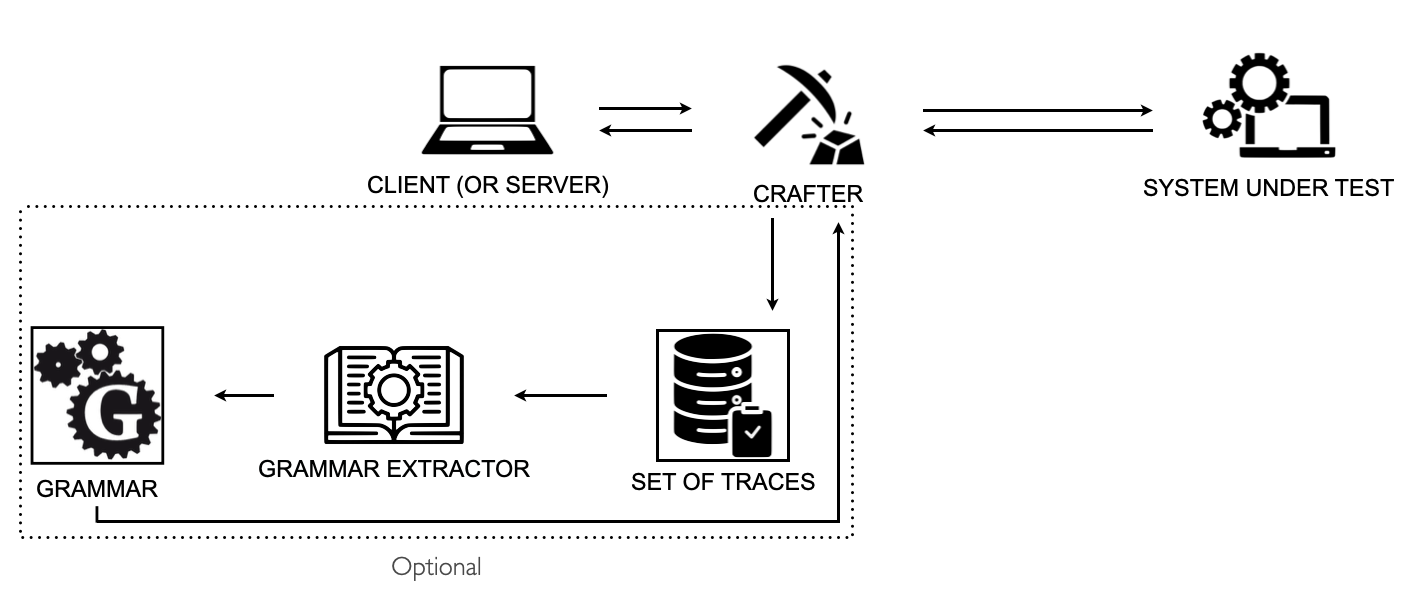}
 \caption{Man-in-the-middle fuzzers}
 \label{fig:man_in_the_middle_fuzzers}
\end{figure}

The final two categories of fuzzers are very different from the five
above:
\begin{enumerate} \setcounter{enumi}{5}
  \item \textbf{\textit{Man-in-the-Middle fuzzers}}: These fuzzers sit
       in the middle between the SUT and a program interacting with it
       and modify messages going to the SUT, as illustrated in
       Fig.~\ref{fig:man_in_the_middle_fuzzers}.  Responses coming
       back from the SUT are left untouched.

       These fuzzers can take a dumb mutational approach to modify the
       messages, but they may leverage a protocol specification
       (automatically inferred or given as input) to modify messages.

 \item \textbf{\textit{Machine learning fuzzers}} These fuzzers use a
      Machine Learning (ML) model trained on a large set of input
      traces. The model outputs slightly different --- hopefully
      malicious --- mutated traces.  Machine learning methods used by
      these fuzzers include Seq2seq and Seq-GAN.

      These fuzzers are similar to the grammar learner fuzzers in that
      they require a set of sample traces as input that is then used
      to infer a model of the input format which is then the basis for
      the fuzzing.  The key difference is that for the grammar learner
      fuzzers this model is a grammar, whereas for these machine
      learning fuzzers the model is an ML model.
\end{enumerate}
%

\begin{figure}
\centering
 \includegraphics[scale=0.30]{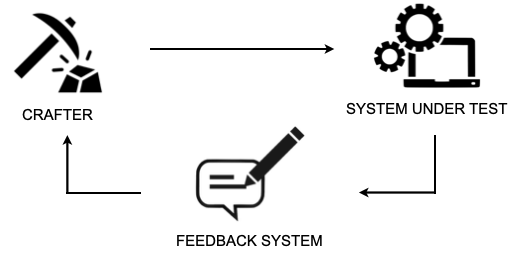}
 \caption{Evolutionary fuzzers }
 \label{fig:evolutionary_fuzzers}
\end{figure}

\begin{figure}
\centering
 \includegraphics[scale=0.30]{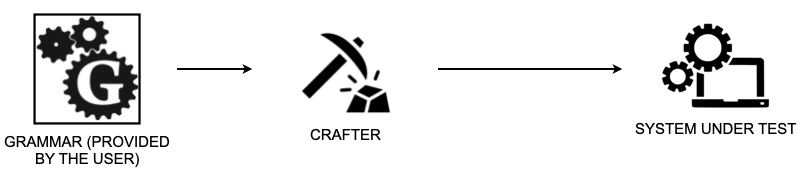}
 \caption{Grammar-based fuzzers }
 \label{fig:grammar_fuzzers}
\end{figure}

\begin{figure}
\centering
 \includegraphics[scale=0.30]{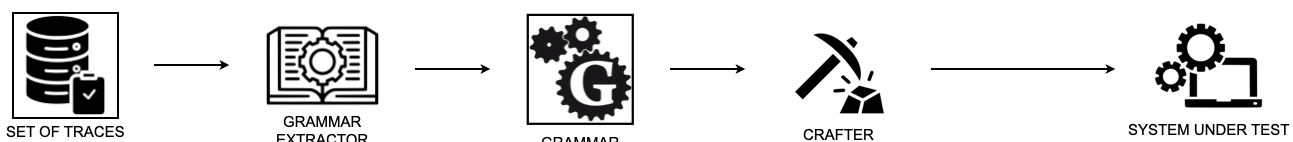}
 \caption{Grammar learner fuzzers}
 \label{fig:grammar_learner_fuzzers}
\end{figure}

\begin{figure}
\centering
 \includegraphics[scale=0.30]{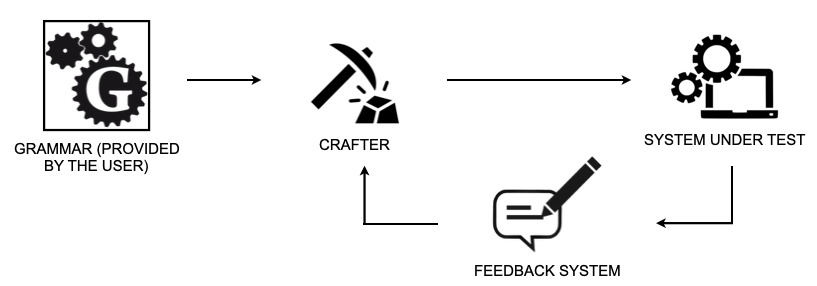}
 \caption{Evolutionary grammar-based fuzzers }
 \label{fig:evolutionary_grammar_fuzzers}
\end{figure}

\begin{figure}
\centering
 \includegraphics[scale=0.25]{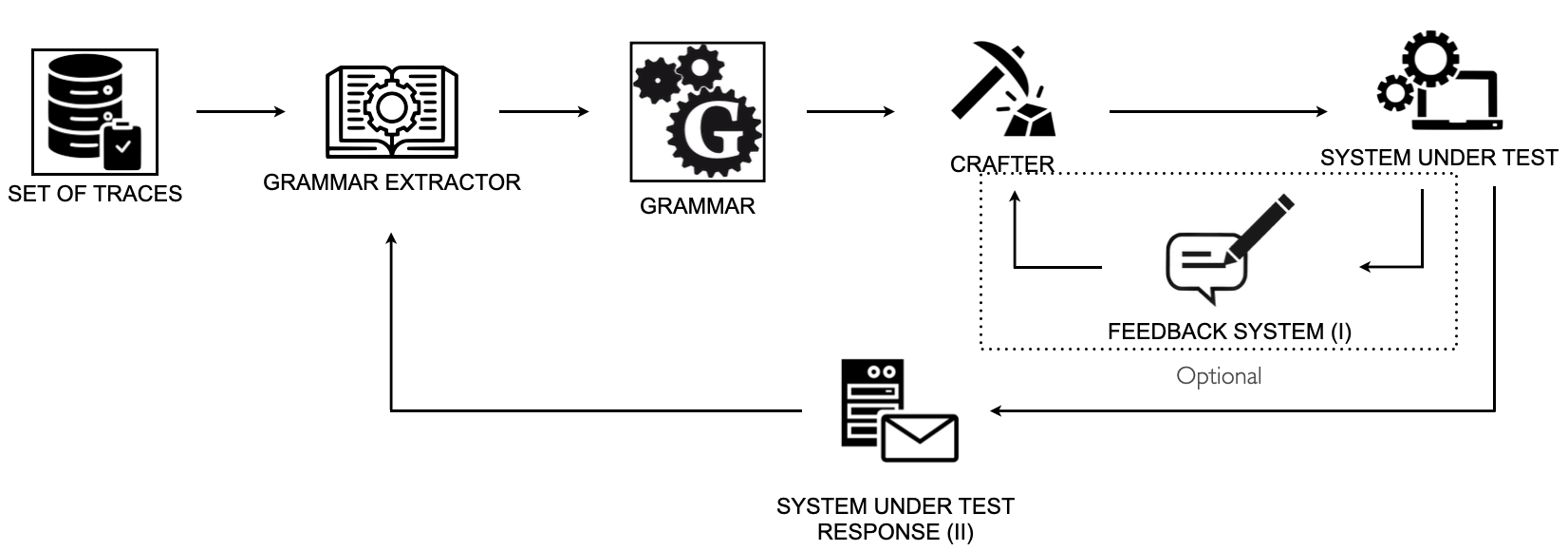}
  \caption{Evolutionary grammar learner fuzzers }
 \label{fig:evolutionary_grammar_learner_fuzzers}
\end{figure}

\begin{figure}
\centering
 \includegraphics[scale=0.30]{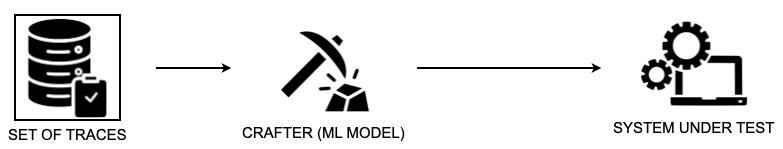}
 \caption{Machine learning fuzzers}
 \label{fig:machine_learning_fuzzers}
\end{figure}

\subsection{Grammar-based fuzzers}\label{sec:grammar_based_fuzzers}

\begin{table}
\centering
\begin{tabular}{|l|l|l|}
\hline
{\textbf{Fuzzer}} & \textbf{Based on}& \textbf{Mutates} \\ \hline
Peach    & - &- Message \\ \hline
SNOOZE  \cite{banks2006}  & - &- Message \\ \hline
PROTOS\cite{kaksonen2001software}  & - &- Message \\ \hline

Sulley   & - &- Message\\ \hline
BooFuzz \cite{pereyda2019boofuzz} & Sulley  &- Message \\ \hline
Fuzzowski \footnote{https://github.com/nccgroup/fuzzowski}  & BooFuzz \cite{pereyda2019boofuzz} &\makecell[l]{- Message\\- Trace}\\ \hline
AspFuzz \cite{aspfuzz}  & -  &\makecell[l]{- Message\\- Trace}\\ \hline
\end{tabular}
\caption{Grammar-based fuzzers}
\label{table:grammar_based_fuzzers}
\end{table}

Table~\ref{table:grammar_based_fuzzers} lists the grammar-based
fuzzers. As shown in Fig.~\ref{fig:grammar_fuzzers}, these fuzzers use
a grammar provided by the user.  In the case of a stateful SUT, this
grammar should describe the syntax of the messages and the protocol
state machine.  For some fuzzers, e.g.\ Peach\footnote{Here we mean the
community edition, available at
\url{https://gitlab.com/gitlab-org/security-products},
which lacks some features of Peach Fuzzer Professional.}
 and SNOOZE \cite{banks2006}, this grammar
is supplied in some XML format.

The obvious downside of these fuzzers is that they require an accurate
grammar. Producing one can be a lot of work and it can be challenging
and error-prone. Not all errors will matter -- or matter equally: if
the grammar is a bit too `loose' this is not much of a problem, but if
the grammar omits interesting parts of the language it may be, as
this would mean that the fuzzer will not explore that part of the
language.  Ideally the documentation of the SUT, or the specification
of the protocol it implements, simply provides a formal grammar that
can be used.  However, this will often not be the case: documentation
or specifications may be unclear or incomplete.  
That the SUT is stateful does not make a difference here, Still, in 
earlier research \cite{Poll2015} we found that documentation is more
likely to include a clear (but informal) specification for the message
format than for the protocol state machine. The protocol state machine
are often -- very poorly -- specified in prose scattered throughout specification documents.

Some grammar-based fuzzers, e.g.\ SNOOZE \cite{banks2006} and Sulley,
come with grammars for some standard protocols, so that for these the
hard work to produce a grammar has already been done for the user, but
for other protocols the user still has to do it themselves.

\subsection{Grammar learner fuzzers}\label{sec:grammar_learner_fuzzers}

\begin{table} 
\centering
\begin{tabular}{|l|l|l|l|}
\hline
\textbf{Fuzzer} &  \textbf{Learns} &  \textbf{Based on}  & \textbf{Input needed} \\
\hline
PULSAR \cite{pulsar}
  & \makecell[l]{- State model\\- Message fields}
  & \makecell[l]{- Passive learning \\(using PRISMA \cite{pulsar})}
  & - Traces
  \\ \hline
GLADE$^+$ \cite{Bastani2017} 
  & \makecell[l]{- Message fields}  
  &  \makecell[l]{- Active learning\\(using GLADE \cite{Bastani2017} )}
  & - Traces 
  \\ \hline
Hsu et al.\cite{hsu2008model} 
  & - State model  
  & \makecell[l]{- Passive learning\\(using partial finite state\\automaton reduction \cite{hsu2008model})}
  & \makecell[l]{- Message field\\specification \\ - Traces}
  \\ \hline
\end{tabular}
\caption{Grammar learner fuzzers}
\label{table:grammar_learner_fuzzers}
\end{table}

Table~\ref{table:grammar_learner_fuzzers} presents the grammar learner
fuzzers. 
These fuzzers operate in two phases:
first, they infer a grammar from a set of collected traces;
then they do the actually fuzzing using that inferred grammar
just like a grammar-based fuzzer would do.
So each of these fuzzers is effectively the composition of two tools:
\begin{enumerate}
    \item a \textit{grammar learner}: a special component with the goal to build a grammar as much as possible similar to the real one
    \item an \textit{actual fuzzer}: in principle any of the grammar-based fuzzers discussed in the previous section. 
\end{enumerate}
All these fuzzers will require a comprehensive and complete set of traces,
as e.g.\ the makers of PULSAR explicitly point out \cite{pulsar}, to give
good fuzzing performance.

For the first phase, the fuzzers in
Table~\ref{table:grammar_learner_fuzzers} not only use different
inference techniques, but also try to infer different aspects of the
input format:
\begin{itemize} 

  \item PULSAR \cite{pulsar} infers both the message format and a
       protocol state machine, passively, from observed traffic.  The 
       learning techniques it uses are the ones developed earlier for
       the PRISMA fuzzer \cite{Krueger2012}. These can also infer rules
       for dependencies between messages, such as increasing sequence
       numbers.

       As the authors note, the approach relies on the completeness of
       the set of observed network traces and will be unable to model
       protocol paths not included in this traffic. 

  \item GLADE \cite{Bastani2017} uses a new active
       learning algorithm for inferring context-free grammars which
       can infer both the message format and the protocol state
       machine. 

       Strictly speaking, GLADE is not a fuzzer, but just a tool for
       inferring a context-free grammar. This inference uses active
       learning, so it does involve some fuzzing of the SUT.  But
       GLADE has been extended to be used as a front-end for a
       grammar-based fuzzer. In
       Table~\ref{table:grammar_learner_fuzzers} we refer to this
       extension as GLADE$^+$ to avoid confusion.

       The algorithm used by GLADE$^+$ is shown to have better precision
       and recall that the active learning algorithms L*
       \cite{Angluin87} and RNPI \cite{Oncina1992} for the case
       studies tried out by the makers \cite{Bastani2017}.  The
       results of GLADE$^+$ are also compared with AFL for some of
       these case studies. However, the case studies are not typical
       stateful protocols but include interpreters for Python, Ruby
       and JavaScript.  As AFL is best at fuzzing binary formats, it
       is maybe not that surprising that GLADE$^+$ beats AFL here.

 \item Unlike PULSAR and GLADE, the fuzzer by Hsu et al.\
      \cite{hsu2008model} cannot infer the message format: it only
      infers a protocol state machine. The tool requires that the
      message format is known; in fact, it needs an (un)parser for the
      message format to be supplied.  The protocol state machine is
      then inferred from observed traffic -- i.e.\ using passive
      learning -- using a new algorithm they introduce.
      Once this state machine is inferred, the SUT can be fuzzed.
      Here a collection of mutation primitives is used, including
      mutations to mutate individual messages and mutations to reorder
      messages in the input trace.

      The first phase of this fuzzer, i.e.\ inferring a protocol state
      machine given a known message format, is very similar to what
      tools like LearnLib \cite{Raffelt2009} do. But it uses passive
      learning, whereas LearnLib uses active learning with a variant
      of L*. Hsu et al.\ report that they also tried active learning
      for this initial phase, using a variant of L*, as they also did
      in earlier work \cite{Shu2008}, but abandoned that approach
      because of 1) the difficulty in constructing concrete messages
      that active learning requires and 2) it being inefficient and
      not learning an accurate model.

\end{itemize}
%

\subsection{Evolutionary fuzzers}\label{sec:evolutionary_fuzzers}

\begin{table}
\centering
\begin{tabular}{|l|l|l|l|l|}
\hline

{\textbf{Fuzzer}} & \textbf{\makecell[l]{Feedback \\ system}}  & \textbf{Based on} & \textbf{Input needed} \\ \hline
  nyx-net \cite{schumilo2021nyx} 
    & - Coverage &   - AFL & \makecell[l]{- Target binary\\- Protocol specification\\- Seed inputs (optional)}
  \\ \hline
  FitM fuzzer \cite{fitm} 
    & - Coverage &   - AFL & \makecell[l]{- Client binary\\- Server binary\\- Seed inputs}
  \\ \hline
  SNPSfuzzer \cite{li2022snpsfuzzer}  
    & \makecell[l]{ - Coverage } &   - AFL & \makecell[l]{- Target binary\\- Seed inputs}
    \\ \hline

  Chen et al. \cite{exploringchen2019} 
  & \makecell[l]{ - Coverage \\ - \BRANCH} &  \makecell[l]{- AFL\\- Manual code\\annotation}  & \makecell[l]{- Target binary\\- Seed inputs}
  \\ \hline
  SGFuzz \cite{ba2022stateful} 
  &\makecell[l]{- Coverage \\ - Variables} &  \makecell[l]{- AFL \\ - Automatic code\\annotation }  & \makecell[l]{- Target binary\\- Seed inputs}
  \\ \hline
  IJON \cite{IJON2020} 
  & \makecell[l]{ - Coverage \\ - Variables} &  \makecell[l]{- AFL\\- Manual code\\annotation}  & \makecell[l]{- Target source code}

  \\ \hline
\end{tabular}
\caption{Evolutionary fuzzers}
\label{table:evolutionary_fuzzers}
\end{table}

Table~\ref{table:evolutionary_fuzzers} presents the evolutionary
fuzzers.  As shown in Fig.~\ref{fig:evolutionary_fuzzers}, these use
feedback to guide the mutation of inputs. This feedback can use
different types of observation, namely the five options listed below
or a combination:
\begin{description}
  \item[F1] \emph{Response}. Some fuzzers use the response of the SUT.  This
    is the only type of observation that can be done black-box.
  \item[F2] \emph{Coverage}. Some fuzzers observe branch coverage in the
    style of AFL, i.e.\ using a bitmap to observe branches taken
    during execution. This is a greybox approach that either requires
    re-compilation to instrument the code or running code in some
    emulator, just like AFL does.
  \item[F3] \emph{Branches}: Some fuzzers observe branch coverage not by
    observing all branches like AFL does, but by observing specific
    branches that are manually marked as interesting to the user. This is
    a white-box approach and requires manual annotation of code by the
    user.
  \item[F4] \emph{Variables}: Some fuzzers observe the value of specific
    program variables. This is a white-box approach and requires
    manual annotation of code by the user.  The idea is that the
    program variables observed record information about the
    protocol state.
  \item[F5] \emph{Memory}: Instead of observing specific individual
    program variables, one fuzzer observes memory segments: it
    takes snapshots of memory areas to see if inputs affect these.
    The idea is that changes in the memory signal change the
    protocol state.  The only fuzzer using this, StateAFL, is not an
    evolutionary fuzzer but one of the more complicated evolutionary
    grammar learner, so it is discussed in
    Section~\ref{sec:evolutionary_grammar_learner_fuzzers}.  
\end{description}
All the evolutionary fuzzers in Table~\ref{table:evolutionary_fuzzers}
are based on AFL, so all of them at least observe branch coverage in
the style of ALF (i.e.\ F2, \emph{Coverage}) to
steer the evolution, but some tools use an additional feedback
mechanism on top of this. 

Regarding F3: the fuzzer by Chen et al.\ allows the user to mark some
specific branches in the code.  The idea is that taking these marked
branches is an indication of the SUT moving to a different protocol
state.  Given that the AFL instrumentation already observes branch
coverage, it is somewhat surprising that additional observation of
selected branches improves the performance of the fuzzer.  The fuzzer
not just observes if these branches are taken in execution, but when
this happens it effectively starts a new AFL session for this specific
state (i.e.\ using a new bitmap for recording branches and creating a
new queue of messages to mutate).  So whereas AFL and all the other
AFL-like evolutionary fuzzers in
Table~\ref{table:evolutionary_fuzzers} maintain a single bitmap to
record which branches have been taken, the fuzzer by Chen et al.\ has
one such bitmap for each of the marked branches.  This allows it to
learn different strategies for generating test cases for different
protocol states.  Intuitively this makes sense: messages in different
stages of a protocol may have different formats, so learning different
mutation strategies, each tailored to a specific protocol state, can
improve the fuzzing.  

Regarding F4: IJON \cite{IJON2020} observes specific program variables
during the fuzzing.  The user has to mark these in the source code.
The idea is that the user marks variables that record
information about the protocol state.  SGFuzz is an improvement of
this: instead of the user having to annotate code to specify which
program variables record interesting state information, the fuzzer
automatically infers which program variables have an enumeration type,
and it assumes that all these program variables record state
information.

As discussed in Section~\ref{sec:terminology}, there are different
ways in which the SUT can record its protocol state. If the protocol
state is recorded in program variables, approach F4 of IJON and
SGFuzz can be expected to work well.  If the program point is used the
protocol state, approach F3 as used by Chen et al.\ might work
better.

All evolutionary fuzzers require initial seeds as input traces to
start fuzzing. The choice of these initial seeds can influence the
performance.  Some fuzzers provide some automation to create
initial seeds: for instance, Nyx-net \cite{schumilo2021nyx} provides
functionality (in the form of a Python library) to generate seeds
messages from PCAP network dumps.  The creators of IJON
\cite{IJON2020} note that in some cases IJON's feedback mechanism
works so good that manually picking good seeds is no longer necessary
to obtain good coverage; in some experiments, they could simply use a
single uninformative seed containing only the character `a'
\cite{IJON2020} .

\subsection{Evolutionary grammar-based fuzzers}\label{sec:evolutionary_grammar_based_fuzzers}

\begin{table} 
\centering
\begin{tabular}{|l|l|l|l|} \hline
  \textbf{Fuzzer} & \textbf{\makecell[l]{Feedback\\ system}} & \textbf{Based on}& \textbf{Inputs needed}  \\ 
  \hline
  RESTler \cite{RESTler} & - Response  & - & \makecell[l]{- State model specification 
  \\- Target source file}
  \\ \hline
  SPFuzz \cite{spfuzz}   & - Coverage  & - AFL & \makecell[l]{
     - Protocol specification
   \\- Target source code
   \\- Initial seeds}
  \\ \hline
  EPF \cite{Helme2021}
    &  - Coverage &  \makecell[l]{- AFL\\ - Fuzzowski} 
    & \makecell[l]{- Protocol specification
                   \\- Target source code
                   \\- PCAP files (as initial seeds)}    
  \\ \hline
  \end{tabular}
\caption{Evolutionary grammar-based fuzzers}
\label{table:evolutionary_grammar_based_fuzzers}
\end{table}

Table~\ref{table:evolutionary_grammar_based_fuzzers} presents the
evolutionary grammar-based fuzzers.  These fuzzers combine the
\textit{grammar-based} and \textit{evolutionary} approaches, as shown
in Fig.~\ref{fig:evolutionary_grammar_fuzzers}:
they require a grammar as the starting point to generate messages
but they also include some feedback to observe the effects
of inputs in an effort to fuzz more intelligently.

RESTler \cite{RESTler} is an open-source fuzzer by Microsoft for
fuzzing REST APIs. It uses a grammar in the form of an
OpenAPI\footnote{\url{https://www.openapis.org}} specification (as can
be produced by Swagger tools) to generate messages but then observes
responses combinations of messages that always lead to the same error.
The fact that RESTful API typically come with grammar in the form of
an OpenAPI spec is a big win: it means we can use a grammar-based
approach but avoid the downside of having to produce a grammar. 

SPFuzz \cite{spfuzz} and EPF \cite{Helme2021} observe branch coverage
in the style of AFL to get information about coverage (i.e.\ F2)),
whereas RESTler only uses the response of the SUT (i.e.\ F1).  .

RESTler does not require any initial seeds provided by the user, as
you would expect of a fuzzer that has a grammar that can be used to
generate inputs.  However, SPFuzz and EPF do require the user to
provide initial seeds.  For EPF these are provided in the form of a
PCAP file.

SPFuzz does not use some standard specification format like
OpenAPI, but it has its own format to describe the protocol grammar
and dependencies. 

These dependencies, like the ones between requests and responses \cite{RESTler},
or the ones between the length
field, the content of the message or the data types \cite{spfuzz,Helme2021}, significantly influence the
quality of the inputs generated by the fuzzer.

\subsection{Evolutionary grammar learner fuzzers}\label{sec:evolutionary_grammar_learner_fuzzers}

\begin{table}[ht]
\centering
\begin{tabular}{|l|l|l|l|l|l|}
\hline
  \textbf{Fuzzer} &\textbf{\makecell[l]{Learns}} &\textbf{Feedback (i)} & \textbf{Feedback (ii)} &\textbf{Inputs needed} &\textbf{Based on} 
  \\ \hline
  AFLNet \cite{aflnet} 
    & - State model & - Coverage & \makecell[l]{- Response} & \makecell[l]{- Target binary\\- Sample traces} & - AFL
  \\ \hline
    FFUZZ \cite{9565673}  & - State model &- Coverage  & - Response & \makecell[l]{- Target binary\\- Sample traces} & - AFLNet 
  \\ \hline
  StateAFL \cite{Natella2021} 
    & - State model & - Coverage  & \makecell[l]{- Memory} & \makecell[l]{- Target binary\\- Sample traces} & - AFLNet 
  \\ \hline
  SGPFuzzer \cite{SGPFuzzer}
    & \makecell[l]{- State model\\- Message fields} & - Coverage & \makecell[l]{- Response}& \makecell[l]{- Target binary\\- PCAP file} & \makecell[l]{- AFL }
  \\ \hline
  LearnLib \cite{Raffelt2009}
     & - State model & N/A     & - Response & 
     \makecell[l]{- Set of messages}  & \makecell[l]{- TTT \cite{isberner2014ttt}}
  \\ \hline
  Doupé et al. \cite{Doupe2012}
    & - State model &\makecell[l]{N/A}  & \makecell[l]{- Response} & No input required
    
    & \makecell[l]{- Web crawling }
  \\ \hline
\end{tabular}
\caption{Evolutionary grammar learner fuzzers}
\label{table:evolutionary_grammar_learner_fuzzers}
\end{table}

\noindent Table~\ref{table:evolutionary_grammar_learner_fuzzers}
presents the evolutionary grammar-learner fuzzers.  This is the most
complex category of fuzzers.  These fuzzers involve a grammar, which
only describes (an approximation of) the protocol state machine.  They
use two forms of evolution, illustrated by the two feedback loops in
Figure~\ref{fig:evolutionary_grammar_learner_fuzzers}:

\begin{enumerate}[(i)]
  \item \emph{Message evolution}: like for the evolutionary fuzzers
       discussed in Section~\ref{sec:evolutionary_fuzzers}, feedback
       from the system is used to mutate traces, using one or several
       of the five types of observation discussed there. 

  \item \emph{State machine evolution}: here feedback from the system
       is used to improve an approximation of the protocol state
       machine.  This comes down to a form of active state machine
       learning.

       For all tools except StateAFL the feedback used here is the
       response from the SUT (i.e.\ F1) or some information extracted
       from that response; for example, for AFLNet it is the response
       code in the response, for EPF it is just information about
       whether the connection was dropped.
       StateAFL observes whether the content of long-lived memory 
       areas has changed (i.e.\ F5).
\end{enumerate} 

LearnLib and the fuzzer by Doup\'e et al.\ are odd ones out in
Table~\ref{table:evolutionary_grammar_learner_fuzzers} in that they
are very limited in the kind of fuzzing they do.  They do not mutate
individual messages but only try combinations of a fixed set of input
messages to infer the state machine. Here LearnLib uses the TTT
algorithm \cite{isberner2014ttt}, an improvement of L*. The fuzzer of
Doup\'e et al.\ uses an ad-hoc algorithm developed for the tool: it is
a fuzzer for web applications, so the response of the SUT is a web
page, and the tool analyses these web pages for similarity in an
attempt to crawl the entire website.


\subsection{Man-in-the-middle fuzzers}\label{sec:man_in_the_middle_fuzzers}

\begin{table}
\centering
\begin{tabular}{|l|l|l|l|}
\hline
\textbf{Fuzzer} & \textbf{Limitations} & \textbf{Uses} & \textbf{Input needed} \\ \hline
AutoFuzz \cite{Gorbunov2010} 
   & \makecell[l]{- Cannot fuzz the message order} 
   & \makecell[l]{Passive\\learning \cite{beddoe2004network} \cite{hsu2008model}} & Live traffic 
   \\ \hline
\makecell[l]{Black-Box Live \\ Protocol Fuzzing \cite{ramsauer2black}}
  & \makecell[l]{- Cannot fuzz the message order\\- User needs to specify\\the fields to fuzz}
  & N/A & Live traffic\\ \hline
SECFuzz \cite{tsankov2012secfuzz}  
  & - Limited fuzzing of message order
  & N/A & Live traffic \\ \hline
\end{tabular}
\caption{Man-in-the-middle fuzzers}
\label{table:man-in-the-middle-fuzzers}
\end{table}

Table~\ref{table:man-in-the-middle-fuzzers} presents the
man-in-the-middle-fuzzers.  As shown in
Fig.~\ref{fig:man_in_the_middle_fuzzers}, these fuzzers sit between
the SUT and another application that interacts with the SUT to
intercept the communication and modify the communication going to the
SUT.  If the SUT is a server then this other application will be a
client.

An fundamental limitation of these fuzzers is that they are only
able to modify the order of the messages in a limited way.
In fact, AutoFuzz \cite{Gorbunov2010} and Black-Box Live Protocol Fuzzing
\cite{ramsauer2black} do not modify the order of messages at all,
so the exploration of the protocol state machine will be very
limited.
SECFuzz \cite{tsankov2012secfuzz} does fuzz the
order of the messages, but only a little bit, namely by inserting
well-formed messages at random positions in the input trace 
(i.e.\ in the sequence of messages sent by the other application).

Even though the overall set-up is the same, the fuzzers use different
techniques:
\begin{itemize}
\item Like the grammar learner fuzzers, AutoFuzz operates in two
     phases.  Prior to the actual fuzzing it starts with a passive
     learning phase to infer an approximation of the protocol state
     machine. For this AutoFuzz uses the same algorithm as Hsu et al.\
     \cite{hsu2008model}, i.e.\ partial finite state automaton
     reduction. So, as for Hsu et al.\, the user has to supply
     implementations of abstraction functions that map concrete
     messages to some abstract alphabet. During the fuzzing AutoFuzz
     then its knowledge of the protocol state machine to guide the
     input selection.
\item Black-Box Live Protocol Fuzzing uses a function to
     generate the message field specification from a PCAP file, but the user is required to choose the
     fields of the messages to fuzz.
\item SECFuzz is able to deal with the cryptography of the protocol. To do that, the client has to share
     with the fuzzer (through a log file) all the information
     necessary for the decryption.
\end{itemize}

\subsection{Machine learning fuzzers}\label{sec:machine_learning_fuzzers}

\begin{table} 
\centering
\begin{tabular}{|l|l|l|}
\hline
\textbf{Fuzzer}  & \textbf{Based on}  & \textbf{Input needed} \\ \hline
GANFuzz \cite{GANFuzz2018}  & seq-gan model & Traces
\\ \hline
\makecell[l]{Machine Learning for\\ Black-box Fuzzing of\\Network Protocols \cite{fan2017machine}} & seq2seq model  & Traces
\\ \hline
SeqFuzzer \cite{8730177}  & seq2seq model & Traces
\\ \hline
\end{tabular}
\caption{Machine learning fuzzers}
\label{table:machine_learning_fuzzers}
\end{table}

Table~\ref{table:machine_learning_fuzzers} presents the machine
learning fuzzers. As shown in Fig.~\ref{fig:machine_learning_fuzzers},
like the grammar learner and the evolutionary grammar learner fuzzers,
these fuzzers require a set of traces that are used as dataset to
train the machine learning model.  Once trained, the machine learning
model is able to output traces that slightly differ from legit,
correct traces.  Likewise the man-in-the-middle fuzzers, the machine
learning fuzzers observe protocol executions that follow the happy
flow.  This cause an unbalanced dataset in favour of the correct traces
and the model's inability to outcomes traces with messages in the
wrong order.

Although these fuzzers use a machine learning model --- trained on
real protocol execution --- to output traces to forward to the SUT,
they employ different strategies. GANFuzz \cite{GANFuzz2018} uses a
generative adversarial network (GAN) and an RNN (recursive neural
network), while the fuzzer by Fan et al.\ \cite{fan2017machine} and
SeqFuzzer \cite{8730177} use seq2seq.  We refer to the review by Wang
et al.\ \cite{wang2020systematic} for a more exhaustive explanation of
fuzzing using machine learning.

\section{Generic fuzzers improvements}

Irrespective of the category of fuzzer, there are some
generic improvements that several fuzzers include.

\paragraph{Pre-processing of raw network traffic}

Many fuzzers take raw network traffic in the formal of a PCAP file as
input and provide some automated pre-processing of that input.  Each
tool implements it in their own way but it includes some common
ingredients, such as chopping up the traces to extract the individual
messages to then clustering similar messages or recognizing specific
fields in the messages.  

\paragraph{Using snapshots}

One factor that makes fuzzing of stateful systems slow is that a
fuzzer often needs to repeat a sequence of inputs to get the SUT in a
particular state, to then start the actual fuzzing in that program
state. To avoid the overhead, some fuzzers
\cite{schumilo2021nyx,fitm,li2022snpsfuzzer} use
snapshots (aka checkpoints) to capture the program state of the SUT at
a particular point in time, to then be able to quickly re-start
fuzzing from that point on.  (The same idea is behind the use of
forking by AFL, where even for stateless SUTs it has been shown to
improve performance off.) This can speed up fuzzing, as the initial
trace to reach some specific state does not have to be repeated,
but taking and re-starting snapshots also introduces overhead, so
in the end it may not be faster.  Depending on the execution platform
there are different snapshotting techniques that can be used.  For
instance, FitM and SNPSFuzzer use CRIU’s userspace snapshots and 
nyx-net uses hypervisor-based snapshot. For
SNSP different snapshotting technologies have been compared
\cite{li2022snpsfuzzer}: CRIU
\footnote{https://github.com/checkpoint-restore/criu}, DMTCP
\cite{ansel2009dmtcp}, BLCR
\footnote{https://github.com/angelos-se/blcr}, and OpenCZ
\footnote{https://openvz.livejournal.com}.

\paragraph{Mutation primitives and heuristics}

Any fuzzer that uses some form of mutation (of individual messages or
of traces) can use a variety of strategies and primitives to do this.
For individual messages this may include random bit-flipping, deleting
some parts of a message or inserting some data.  For traces as opposed
to individual messages) interesting mutation primitives are of course
removal, insertion, or repetition of messages.  

The fuzzers we discussed come with variety of primitives for all this.
Some offer possibilities for the user to provide their own custom
mutators.  We have not gone into the details of this, as the focus was
on understanding the overall approach.  Some fuzzers, notably
SNOOZE, PROTOS, SPFuzz, SGPFuzzer, and the fuzzer by Hsu et al.\
\cite{hsu2008model}, provide more advanced heuristics and tricks for
mutations than some of the others.  For example, SNOOZE can provide
mutations to try out SQL or command injection or use specific numbers
to test boundary conditions.  SPFuzz distinguishes different types of
data inside messages (e.g.\ headers vs payloads) to then use different
mutation strategies for specific types of data.  
In practice it may of course make a big difference for a particular
case study which mutation primitives or heuristics are used. 

\begin{figure}
\centering
\includegraphics[scale=0.45]{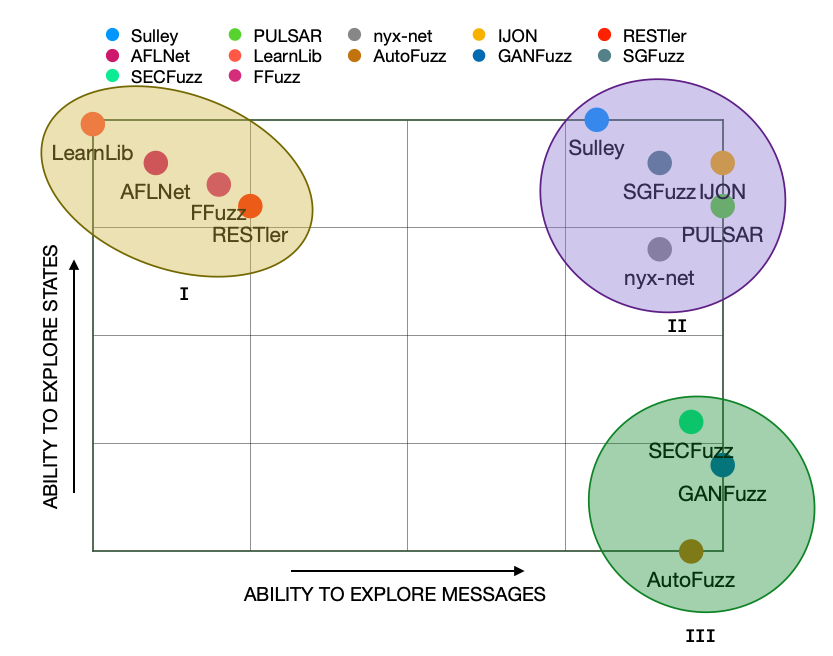}
\caption{Cluster of fuzzers}
\label{fig:graph-comparison}
\end{figure}

\section{Conclusions}

It took us quite some effort to disentangle the ways that various
fuzzers for stateful systems work and arrive at the classification we
presented.  It seems like every fuzzer picks another set of building
blocks, combines them in its own way, and then adds some ad-hoc
heuristics and possibly performance optimisations.  New fuzzers are
typically evaluated on some case studies and then compared with some
other fuzzers, but it is hard to draw broader conclusions that then go
beyond a particular case study or a particular pair of fuzzers.  This
underlines the importance of initiatives such as ProFuzzBench
\cite{Natella2021} for bench-marking stateful fuzzing approaches.
Benchmarking has also been pointed out as a challenge for fuzzers in
general, not just for stateful fuzzing \cite{Boehme2021}. 

We have noted some apparent contradictory observations -- though this
may simply be because researchers looked at different case studies.
For instance, Shu et al.\ \cite{Shu2008} abandoned the use active
learning of protocol state machine using L* (or its variants) because
they found it too slow and inaccurate, while in other research this
has proved to be very useful in finding security flaws
\cite{deRuiter2015}. 

It is not surprising that the performance of fuzzer may depend heavily
on the case study.  When fuzzing a stateful system there is a
trade-off between a) trying out many variations of individual messages
and b) trying out many different \emph{sequences} of messages.  The
complexity of an application (and hence the likely problem spots)
application may more in the message format or more in the protocol
state machine; a corresponding strategy when fuzzing, focusing more
on a) or on b), is then most likely to find bugs.
Very broadly we can make a rough distinction into three classes of
tools, 
illustrated in Fig.~\ref{fig:graph-comparison}:
I) fuzzers that are very good at aggressively exploring
the protocol state machine but poor at trying out variations of
messages;
III) fuzzers that are good at trying out variations in messages
but poor at exploring the protocol state machine;
and
II) fuzzers which try to explore both the protocol state machine
and the format of individual messages.

There is some relation between this classification and the seven
categories we have described.  For instance, the man-in-the-middle
fuzzers and the machine learning fuzzers are in class III, as they do
not explore the protocol state machine and mainly (or even
exclusively) stick to message sequences observed in real
communications between client and server. The grammar-based fuzzers
can deal quite well with both dimensions of fuzzing so are in class
II.  Evolutionary-based fuzzers that try to infer the protocol state
machine (typically using the response of the SUT as feedback mechanism)
are good at exploring the protocol state space, but may lack mutation
primitives or observation mechanisms to aggressively explore the
message formats.  LearnLib is an extreme instance of class I as it
\emph{only} fuzzes the message order.

The exact positioning of tools in Figure~\ref{fig:graph-comparison} is
not based on experimental data, but more informally based on the
general characteristics of the tools, so should be taken with a pinch
of salt. Also, for tools that require grammars as input or manual code
annotation a lot will depend on the quality of these.

It may seem like fuzzers of type II are the best of both worlds, but
given the rapid state space explosion when we fuzz both individual
messages and sequences of messages this need not be the case: Using a
fuzzer of type I and a fuzzer of type III to explore different aspects
may be more effective than using one fuzzer of type II that tries to
do both.  For fuzzing of non-stateful systems it has already
demonstrated that using a combination of tools may be the optimal
approach, especially if these tools can exchange information
\cite{EnFuzz};  we expect that this will be even more so for stateful
systems.

\smallskip

By providing insight into the components used in various fuzzing
approaches, our research suggests several interesting directions for
future research.  One direction is in trying our new combinations of
approaches and components,  for example, using LearnLib as a
pre-processing phase may be useful get a good initial approximation of
the protocol state machine, or using the SUT response as feedback in
man-in-the-middle fuzzers to build a more accurate protocol state
model.  Some of the performance optimisations implemented by specific
fuzzers (e.g.\ the use of snapshots) can be applied to a broader set
of fuzzers.  Another direction is in more systematic, empirical
comparison: having identified that some tools use the same overall
approach but a different algorithms for some sub-component allows a
more systematic comparison, where we just observe the effect of
changing this one sub-component.

\bibliographystyle{ACM-Reference-Format}
\bibliography{bib}

\end{document}